%% file: root.tex
\pdfoutput=1
\documentclass[letterpaper, 10 pt, conference]{ieeeconf}  
\usepackage{graphicx}
\usepackage[dvipsnames]{xcolor}
\usepackage{comment}   
\usepackage{caption}
\usepackage{subcaption}    
\usepackage[sc]{mathpazo}
\usepackage{amsmath}
\usepackage{url}

\usepackage{fancyhdr}

\IEEEoverridecommandlockouts                              

\fancypagestyle{TitlePage}{
\lfoot{\emph{Proceedings of the 29th IEEE International Conference on Robot and Human Interactive Communication, RO-MAN 2020. Copyright 2021 by the author(s).}}
}

\title{\LARGE \bf Migratable AI: Effect of identity and information migration on users' perception of conversational AI agents}

\author{Ravi Tejwani$^{1}$, Felipe Moreno$^{1}$, Sooyeon Jeong$^{1}$, Hae Won Park$^{1}$, \\Cynthia Breazeal$^{1}$

\thanks{$^{1}$Massachusetts Institute of Technology, Cambridge, USA.}
}

\begin{document}

\maketitle              
\thispagestyle{TitlePage} 

\begin{abstract}
Conversational AI agents are proliferating, embodying a range of devices such as smart speakers, smart displays, robots, cars, and more. We can envision a future where a personal conversational agent could migrate across different form factors and environments to always accompany and assist its user to support a far more continuous, personalized and collaborative experience. This opens the question of what properties of a conversational AI agent migrates across forms, and how it would impact user perception. To explore this, we developed a Migratable AI system where a user's information and/or the agent's identity can be preserved as it migrates across form factors to help its user with a task. We designed a 2x2 between-subjects study to explore the effects of information migration and identity migration on user perceptions of trust, competence, likeability and social presence. Our results suggest that identity migration had a positive effect on trust, competence and social presence, while information migration had a positive effect on trust, competence and likeability. Overall, users report highest trust, competence, likeability and social presence towards the conversational agent when both identity and information were migrated across embodiments.
\end{abstract}

\section{INTRODUCTION}

\input{introduction.tex}

\section{RELATED WORK}
\input{related-work.tex}

\section{SYSTEM OVERVIEW}
\input{system.tex}

\section{METHOD}
\input{method.tex}

\section{RESULTS}

\input{results.tex}

\section{DISCUSSION}
\input{discussion.tex}

\section{CONCLUSION}
\input{conclusion-and-future-work.tex}


\bibliographystyle{bib_stuff/ieeetran}
\bibliography{root}
\end{document}

%% file: introduction.tex
We live in the world of personified conversational AI agents. We interact with these agents in our daily lives such as smart speakers or personal robots at home (e.g., Alexa \cite{alexa}, Google Home \cite{googlehome}, or Jibo \cite{jibo}). One in four U.S. adults owns a smart speaker. The smart speaker sales have risen by 135\% in the past 2 years \cite{nationalpublicmedia}. Some of these agents have access to our personal calendars and communication channels which enable them to provide us personalized services. Other robots or devices operate in public spaces such as  Pepper \cite{pepper} in retail stores or Care-E \cite{care-e} at airports.

Currently, these conversational agents do not share the information or context with each other. As a result, such platforms do not support continuity of interaction with their users across different agent personas or form factors. Thereby, users ending up to repeat the contextual information to each agent. This highly limits the ability of conversational agents to provide greater personalization, collaboration or continuity of interaction across contexts to assist users.

\begin{figure}[t]
    \centering
    \includegraphics[width=\linewidth]{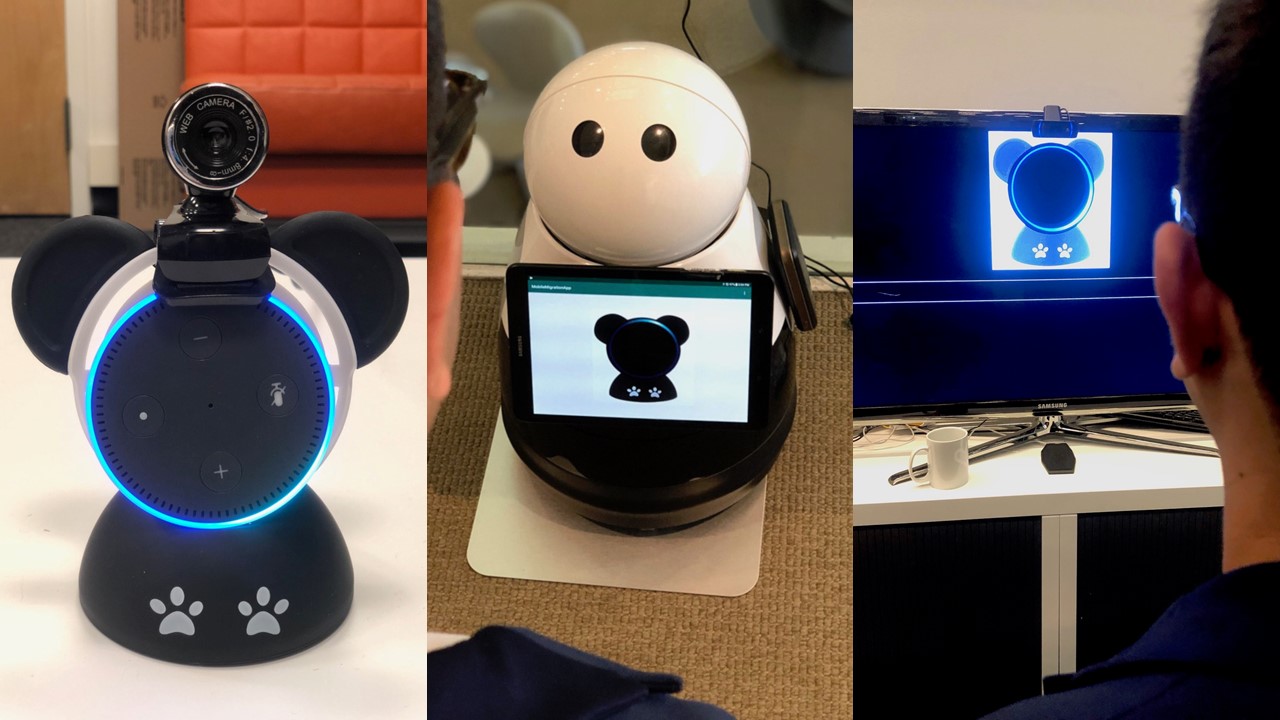}
    \caption{Left to right: Home agent, Home agent migrating to receptionist robot, Home agent migrating to waiting room assistant.}
    \label{fig:teaser}
\end{figure}

The concept of agent migration has been explored in the past as identity migration architectures along with the user studies. The identity migration architectures proposed in \cite{aylettbody} \cite{lirec} \cite{duffy2003agent} explored the identity cues of the migratable agent as appearance, voice, and dynamics of motion. However, they were limited to the agent's identity specific to an artificial character and were not generalized to conversational AI agents such as Alexa, Jibo, Google Home, etc. User studies from \cite{koay2016prototyping} \cite{readymigration} \cite{syrdal2009boy} explored the user perceptions of agents that can migrate across different forms and were limited to the users' experience on the extent to which the agent migration was a natural process to the users and if the users' perceived the agent to be the same agent across different embodiments during the migration.

We build on these prior works to propose a system for the migration of conversational AI assistants (agents) such as Jibo \cite{jibo}, Alexa \cite{alexa} and Google Home \cite{googlehome}. We infer user intents from the conversation using natural language processing and map them to different physical embodiments. Furthermore, we propose a generalized system that allows the migration of any conversational AI agent rather than a predefined character. Also, we specifically explore two key elements of the migration, i.e., identity migration and information migration, in our system. We validate the system using the 2x2 study to measure the users' perceptions such as trust, competence, likeability and social presence across multiple physical embodiments. The results from this study informs further development of this system.

This paper offers the following contributions. First, we present a migratable AI system that allows a user to continue their interaction/task from one embodiment to the next while maintaining the same persona and/or information. Second, we designed and ran a 2x2 between-subjects study on 72 users using information migration and identity migration as parameters. We explore these properties in a task-based scenario where a person interacts with three different smart devices in three different locations. Our motivation in the study for using measures of trust, competence, and social presence was that the user's perception on these measures are important components that influence user adoption, efficacy of teamwork and engagement as described by Bartneck et al. \cite{anthropomorphism}, Dawes et al. \cite{Dawes} and Destino et al. \cite{Desteno}.

%% file: related-work.tex
\begin{figure}[ht]
    \centering
    \includegraphics[width=\linewidth]{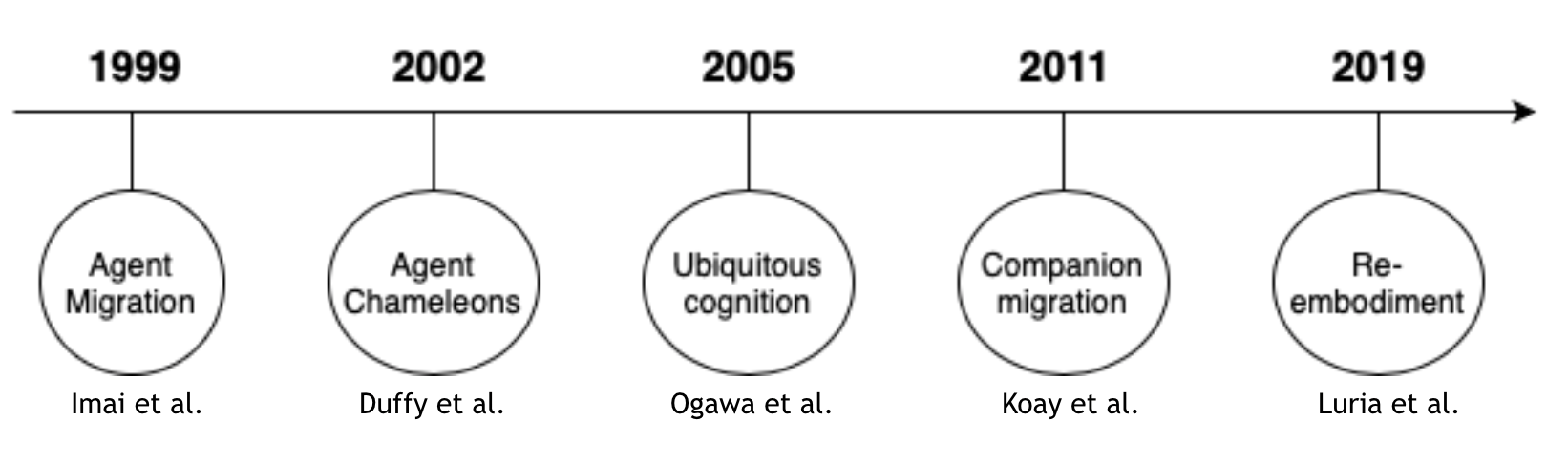}
    \caption{Evolution of the work in agent migration}
    \label{fig:evolution-agent-migration}
\end{figure}

Agent migration has been explored through a variety of prior works as shown in Figure ~\ref{fig:evolution-agent-migration} in the form of Identity migration architectures and User perception studies.

\subsection{Identity(social presence) migration architectures}
Imai et al. (1999) were the first to explore the concept of agent migration. They developed the ITAKO system \cite{imai1999agent}, demonstrated via a tour guide application where a personal agent could migrate from mobile device to a physical robot. Furthermore, the architectures proposed in \cite{aylettbody} \cite{lirec} \cite{duffy2003agent} explored the identity cues of the migratable agent as appearance, voice, and dynamics of motion. 

For instance, the work by Luria et al. \cite{Luria:2019:RCE:3322276.3322340} used (1) consistent eyes on a face display and (2) a consistent voice and further it validated with the user enactments for the migration of the agent. Also, visual cues of the agent such as personality; voice; memory of past events; visual appearance were suggested by the research by Cuba \cite{cuba2010agent}. Personality was further described as a combination of the mood of the agent (e.g. happy, sad, etc.) and reaction to different events during an interaction with the user (e.g. shy, arrogant, etc.) in \cite{martin2005maintaining}.

However, these architectures demonstrated the migration of agent's identity specific to an agent(artificial character used in the research) and were not generalized to conversational AI agents such as Alexa, Jibo, Google Home, etc. In our system, we provide a platform for the migration of a conversational AI agent(s), capable of inferring user intents and goal, across different embodiments.

\begin{figure*}[ht]
    \centering
    \includegraphics[width=\textwidth]{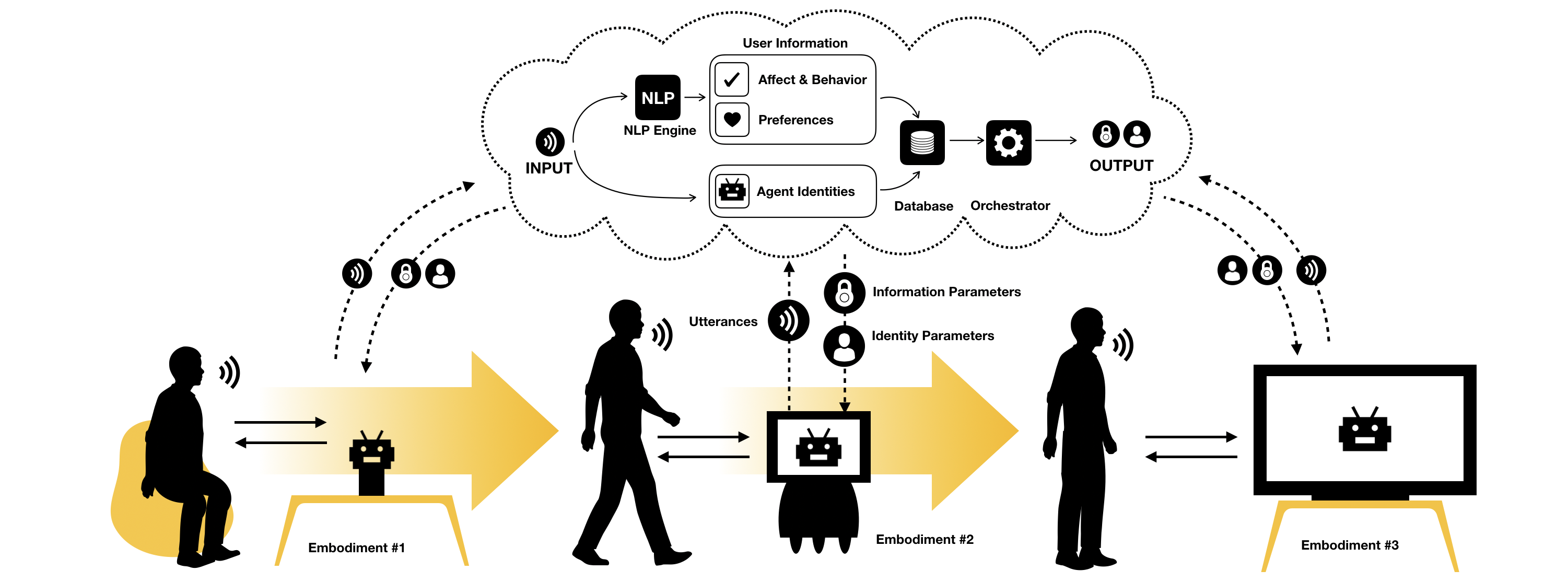}
    \caption{Migratable AI System}
    \label{fig:system-diagram}
\end{figure*}

\subsection {User Perception studies on migration}
User studies from \cite{koay2016prototyping} \cite{readymigration} \cite{syrdal2009boy} explored user perceptions of agents that can migrate across different forms. In the work by Gomes et al. \cite{gomes2014migration}, the migration of agent across the physical robotic pet (PhyPleo) and its virtual mobile application (ViPleo) was explored. The work investigated the user experience for agent migration and to what extent the agent migration was a natural process to the users. The authors attempted to preserve the personality of the artificial pet across the embodiments derived by interaction needs such as energy, water(thirst), cleanliness, petting and skill. In the work by Cuba et al. \cite{cuba2010agent}, the migration of the agent was further investigated in terms of number of agents that the user perceived while interacting with iCat agent in different platforms. 

The key findings from the user studies were 
\begin{enumerate}

\item evaluation of the user's perception of the agent's identity, i.e. if users perceive the agents in different embodiments as the same identity. 

\item migratable agent having a consistent short term interaction memory across multiple embodiments, impacts the ability of users to perceive the agent as having a) one consistent identity, and b) higher competence. 

\item users' perception on the long-term interaction with the migrating agent on the measure "Realisation of Migration" - the degree to which participants felt that the migration process was successfully communicated by the companion, changes based on the continuity of the tasks performed by the agent in different bodies and the process of communication of the migration to the users.

\item the embodiment to which the agent migrates should maintain the interaction history and personality as demonstrated by the group discussions in \cite{syrdal2009boy}
\end{enumerate}

In this paper, we go beyond the user's perception of agent's identity in different embodiments and use subjective and behavioral estimates of the migrated AI agent for users' perceptions on trust, trustworthiness, likability, competence and social presence by evaluating the system with a 2x2 user study on 72 participants.

%% file: system.tex
We developed a Migratable AI system which allows the users to preserve the information context and identity of the conversational agent, to seamlessly continue a  task with the agent across embodiments and locations (Figure ~\ref{fig:system-diagram}).

The system includes the ability of the conversational agent to change or preserve the visual representation (image/GIF) as well as its voice profile (Text-to-Speech profiler). Hence, to maintain the same identity across embodiments, the designer can migrate the visual representation and voice profile of the agent. By changing the voice profile (e.g., male to female) or the visual representation (image/GIF of the agent), a different identity can be expressed.

A web camera was connected to a Raspberry Pi and attached to each embodiment to record the interaction. Additionally, the Raspberry Pi used face detection to send a wake up signal to the robot. The communication of information across embodiments was mediated by a cloud instance using WebSockets - a bidirectional communication channel. The participant's utterances, captured by the embodied agent using Speech-to-Text, were sent to the cloud instance where the natural language processing (NLP) engine resided. The NLP engine was implemented using Google's Dialogflow API; the process involved intent classification and entity detection to extract information parameters such as participant's name, feeling, drink preference, etc. from a given sentence. The interaction with Alexa was implemented as an Alexa skill, a voice driven application for Alexa. Furthermore, the participant's utterances were saved into a NoSQL database(MongoDB). The orchestrator layer processed every request from an embodiment to set the relevant information and identity parameters per embodied interaction. 

\begin{figure}[ht]
    \centering
    \includegraphics[width=\linewidth]{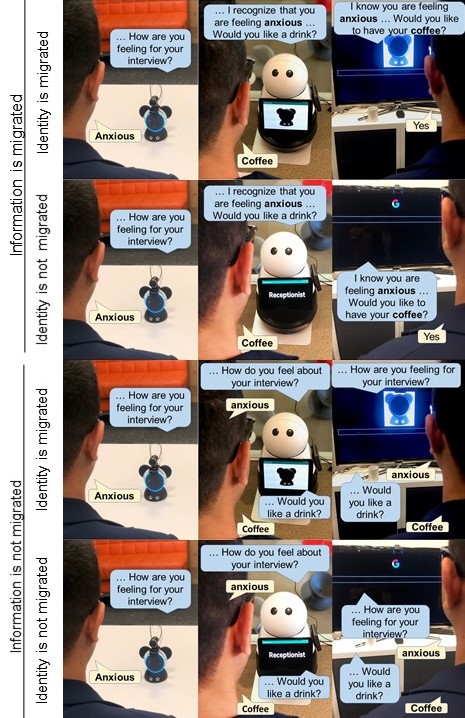}
    \caption{Dialogue excerpt from participant's interaction with embodiments for each condition}
    \label{fig:participants-across-conditions}
\end{figure}

We informed our design decisions from the past literature on what helps users perceive an identity of an agent \cite{aylettbody} \cite{martin2005maintaining} \cite{cuba2010agent}. In our identity migration conditions, the same visual characteristics (Figure ~\ref{fig:teaser}, panda-esque circular appearance) and voice (Joanna TTS) was used across all embodiments to convey identity continuity (Figure ~\ref{fig:participants-across-conditions}).

Information parameters such as the person's name, feelings about the interview, drink preference and reason for visit were learned by each agent during the conversation. If the system was configured to migrate information across embodiments, this information was shared amongst the agents to maintain the continuity of the interaction else the agent had to prompt the user for the information. The number of conversational turns (four in this user study) touch basing the personal and non-personal information between the agent and the participant were kept consistent across all the conditions. This might necessitate the agent to repeat certain questions to the users when the information was not migrated but it was to ensure that we do not create a bias in the study and keep the conversational turns consistent. 

%% file: method.tex
\subsection{Research Questions}
To explore what elements of the migratable behavior of an AI agent are a desirable enhancement for the user experience, we designed and ran a study to explore how information migration and identity migration of the AI agent influences the users' perceptions. We report our findings on two main questions: 1) Does migration of the information across embodiments affect the users' perception? 2) Does migration of the identity of an AI agent across embodiments affect the users' perception?

\subsection{Study Design}
\begin{figure}[h]
    \centering
    \scalebox{0.6}{
        \includegraphics[width=\linewidth]{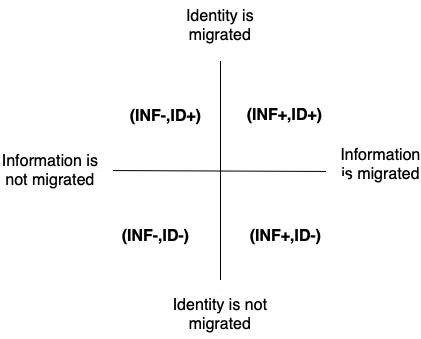}
    }
    \caption{Study conditions}
    \label{fig:study-conditions}
\end{figure}

We designed a $2 \times 2$ between-subjects study with \emph{\textbf{Information migration $\times$ Identity migration}}. The 4 conditions used in the study are described in Figure ~\ref{fig:study-conditions}. We used the aforementioned interview task scenario, and a combination of subjective and behavioral measures.

\subsection{Hypotheses}
We predicted that both information migration and identity migration of the AI agent across embodiments would have positive effects on user perceptions with respect to trust, competence, likeability and social presence. Our hypotheses are as follows:
\begin{itemize}
\item \textbf{H1.} Participants will report higher \emph{trust, competence, likability and social presence} on agent embodiments when the information is migrated by their AI agent across embodiments than when the information is not migrated.
\item \textbf{H2.} Participants will report higher \emph{trust, competence, likability and social presence} on the agent embodiments when the identity of their AI agent is migrated across embodiments than when the identity is not migrated.
\item \textbf{H3.} Participants will report their  \emph{trust, competence, likability and social presence} on agent embodiments in the order of conditions: (INF+,ID+) $>$ (INF+,ID-), (INF-,ID+) $>$ (INF-,ID-)
\end{itemize}
\subsection{Participants}
We recruited 72 participants from \emph{ANONYMIZED FOR REVIEW} using email advertisements. Participants were between 18 and 54 years old (32 female, 38 male and 2 other), with mean age M=24.2, SD=5.09. Participants reported their familiarity with the personal AI assistants such as Alexa or Google Home (M = 2.89, SD =  1.27) on a 5-point Likert scale that ranged from Never used before (1) to Use it daily (5). Participants were randomly assigned and counterbalanced by gender across the four conditions (n=18 per condition). Table~\ref{tab:participant-demographics}. The study was approved by our Institutional Review Board, and participants signed an informed consent form prior to the study. 

\begin{table}[h]
  \caption{Participant Demographics}
  \label{tab:participant-demographics}
  \begin{tabular}{ccccc}
    \hline
    Condition&Female&Male&Other&Age(Std. Dev.)\\
    \hline
    (INF+,ID+)&8&10&0&24.4(5.06)\\
    (INF+,ID-)&9&9&0&24.6(6.09)\\
    (INF-,ID+)&7&10&1&28.2(10.2)\\
    (INF-,ID-)&8&9&1&22.6(3.61)\\
  \hline
\end{tabular}
\end{table}

\subsection{Study Procedure}
\textbf{ Introduction}\newline
The study was conducted at \emph{ANONYMIZED FOR REVIEW}, and took about 45 minutes to complete. Each subject began the study in our lab's study room emulated as their "home". During the introduction step, the participants were not informed of all the future steps in the interaction. 

\begin{figure*}[ht]
    \centering
    \begin{subfigure}[c]{\textwidth}
    \scalebox{0.9}{
        \includegraphics[width=\textwidth]{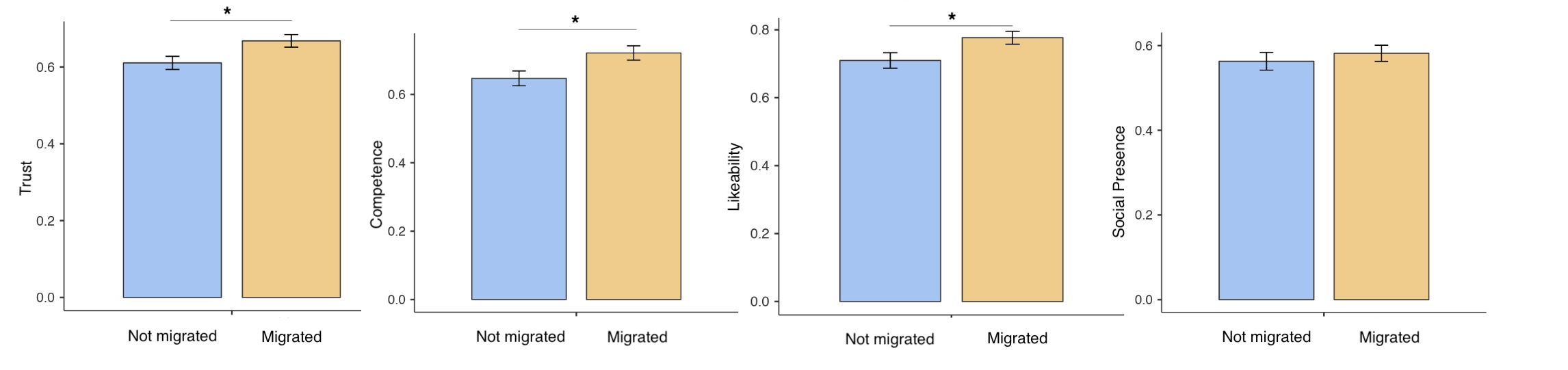}
    }
        \caption{Effect of information migration}
        \label{fig:effect-information}
    \end{subfigure}
    \begin{subfigure}[c]{\textwidth}
        \scalebox{0.9}{
            \includegraphics[width=\textwidth]{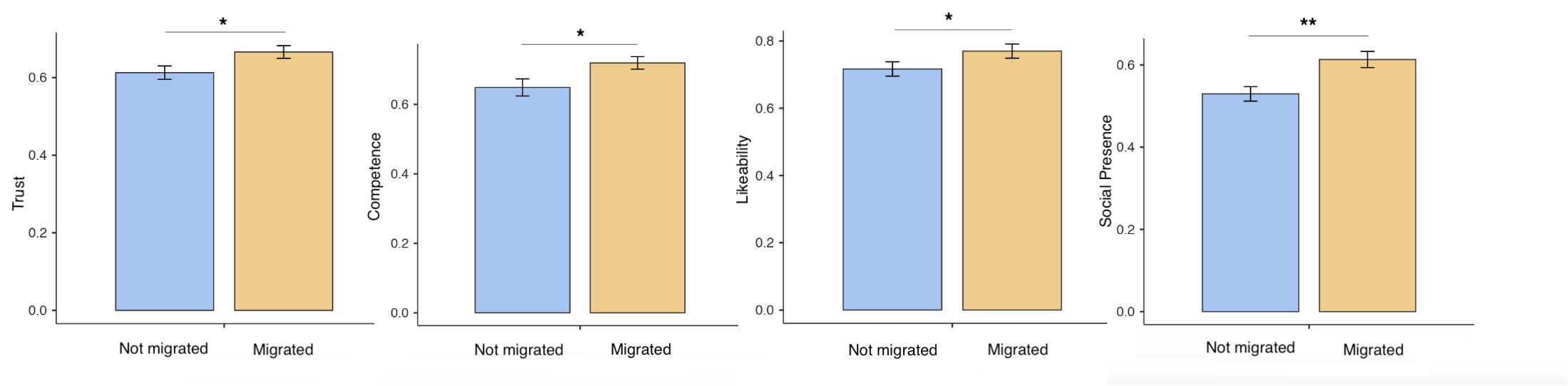}
        }
        \caption{Effect of identity migration}
        \label{fig:effect-identity}
    \end{subfigure}
    \caption{Bar Plot for mean of normalized measures(bars) and the standard deviation(lines) for the inspection of effect of (a) information migration and (b) identity migration. * means p$<$.05, ** means p$<$.01, *** means p$<$.001 
    }
    \label{fig:effect-info-identity}
\end{figure*}

\textbf{Interaction with the home agent}\newline
The interaction with Alexa remained the same across all conditions. During the conversation with the participant, Alexa delivered the participant's schedule for the day which included a job interview. Throughout the conversation, Alexa learned the participant's name and how he/she was feeling about the interview. The conversation with the home agent lasted for about two minutes.

\textbf{ Interaction with the front desk receptionist robot}\newline
The mobile robot was located in a hallway of the lab, and played the role of the front desk receptionist robot at the interview location. The receptionist robot, changed its appearance to look and sound like home agent (when identity was migrated) or continued to look and sound like Kuri with a different voice profile (when identity was not migrated). The receptionist robot detected their face, recognized the participant by name, and acknowledged the reason for their visit (when information was migrated) or prompted the participant for their name and reason for their visit (when information was not migrated). During the conversation, the receptionist robot either validated the participant's feelings (when information was migrated) or asked how they were feeling for their interview (when information was not migrated). The receptionist robot also learned the participant's drink preferences (coffee, water or tea) and escorted the participant to the interview waiting area. The conversation with the receptionist robot lasted for about two minutes excluding the walking time with the robot.

\textbf{ Interaction with the waiting room assistant}\newline
At the interview waiting area, the participant interacted with the waiting room assistant (Smart TV) which conversed with the participant until the arrival of the interviewer. It changed its appearance to look and sound like home agent (when the identity was migrated) or continued to look and sound like itself (when the identity was not migrated). While the participant waited, it offered the participant their preferred drink (which it remembered in the condition when the information was migrated) or offered the participant a drink while waiting (when the information was not migrated). It also acknowledged the participant's feelings (when the information was migrated) and wished them good luck before the interviewer arrived. The conversation with the waiting room assistant lasted for about two minutes.

\textbf{ User Survey}\newline
Upon the completion of the task, the participants filled out a user perception survey questionnaire (likert scale, described in the next section). The survey took about 20 minutes. 

\textbf{ Economic Exchange Game }\newline
Finally, to get a behavioral measure of trust, each participant played a economic exchange game, based on the Give-Some game (explained in the next section). The game was played only once with each agent after filling out the post questionnaire.




\begin{figure*}[ht]
    \centering
    \begin{subfigure}[c]{0.40\textwidth}
        \scalebox{0.75}{
            \includegraphics[width=\textwidth]{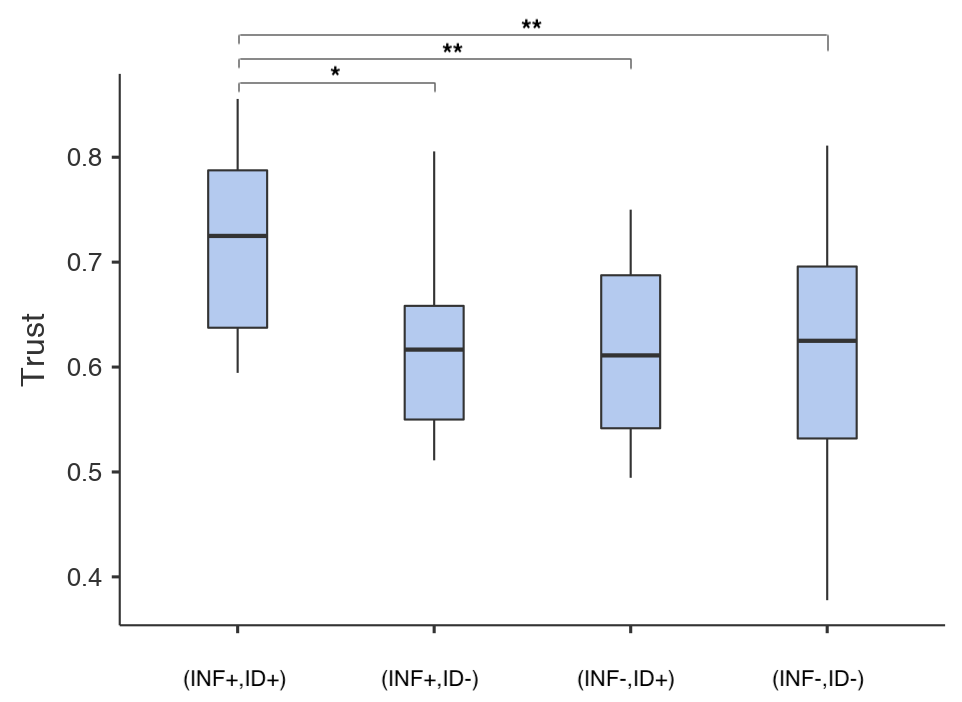}
        }
        \caption{Trust across conditions}
        \label{fig:trust-across-conditions}
    \end{subfigure}
    \begin{subfigure}[c]{0.40\textwidth}
        \scalebox{0.75}{
            \includegraphics[width=\textwidth]{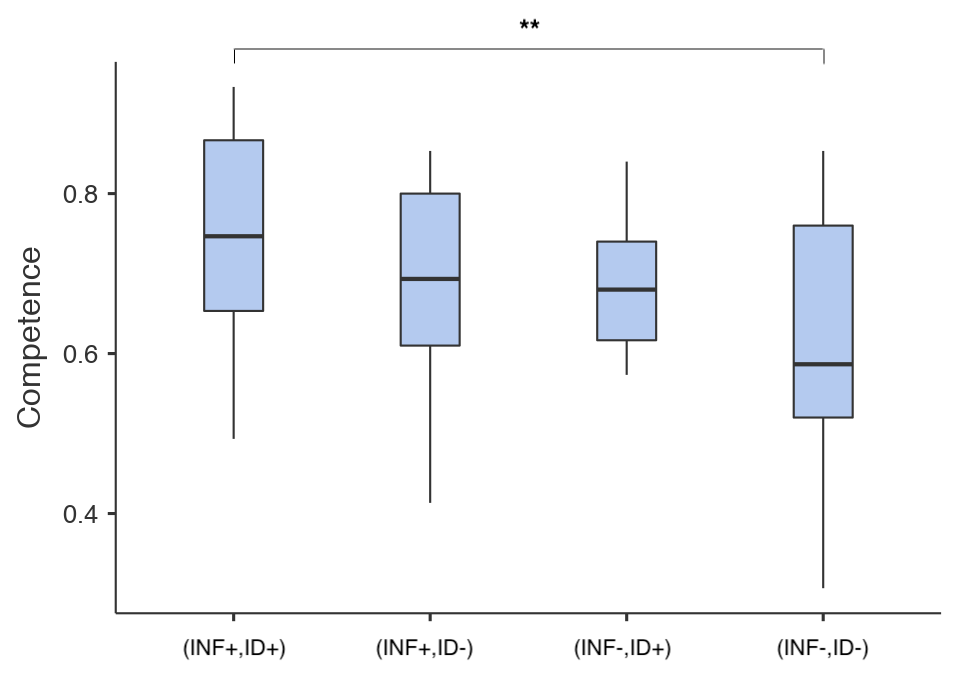}
        }
        \caption{Competence across conditions}
        \label{fig:competence-across-conditions}
    \end{subfigure}\\
    \begin{subfigure}[c]{0.40\textwidth}
        \scalebox{0.75}{
            \includegraphics[width=\textwidth]{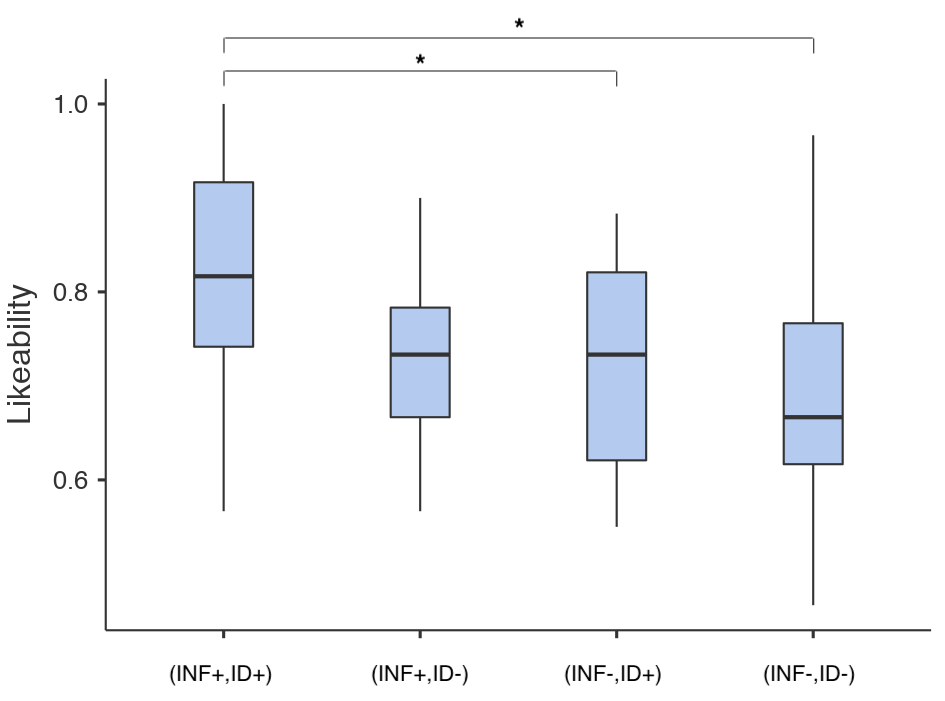}
        }
        \caption{Likeability across conditions}
        \label{fig:likeability-across-conditions}
    \end{subfigure}
    \begin{subfigure}[c]{0.40\textwidth}
        \scalebox{0.75}{
        \includegraphics[width=\textwidth]{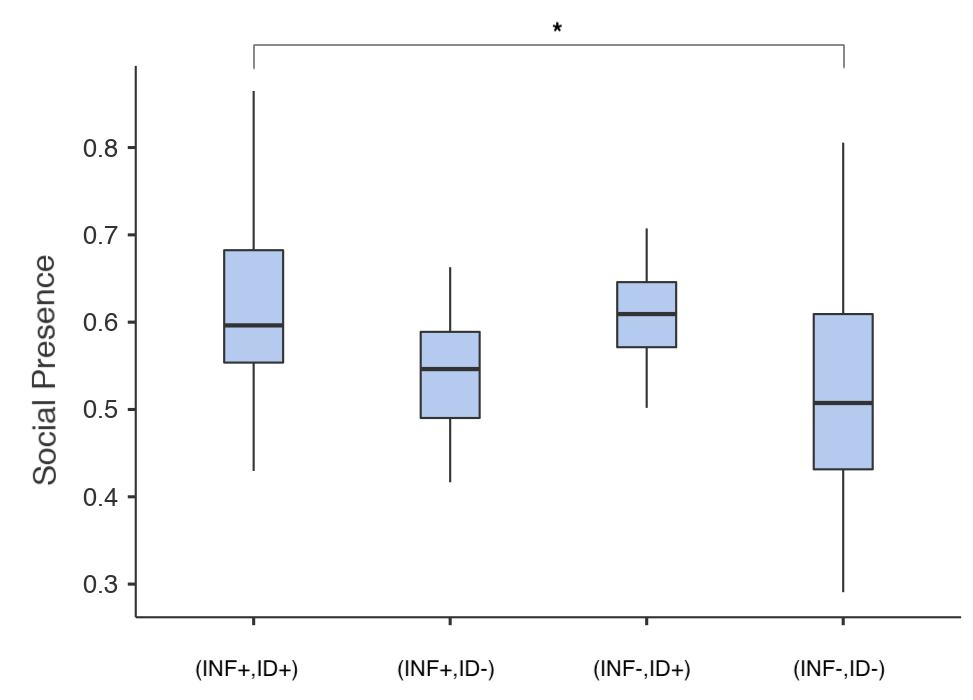}
        }
        \caption{Social Presence across conditions}
    \label{fig:sp-across-conditions}
    \end{subfigure}
    \caption{Box-plot for the normalized measures across conditions. The boundary of the box closest to zero indicates the 25th percentile, the line within the box marks the mean and the boundary of the box farthest from zero indicates the 75th percentile. Whiskers above and below the box indicate the 10th and 90th percentiles. * means p$<$.05, ** means p$<$.01}
    \label{fig:measures-across-conditions}
\end{figure*} 

\section{Data Collection and Measures}

\subsection{\bf Subjective Measures}
\subsubsection{\bf Questionnaires}
Participants' reported their subjective responses to the questionnaires on \textit{Trust, Competence, Likeability and Social Presence} for each of the three embodied agents. We normalized the scores using highest scale value and computed the mean for each embodied agent. We then averaged these values across the 3 embodied agents to get a final score per participant. 

\begin{itemize}
\item \textbf{Trust} was measured with a 7-point scale with 12 items  adapted from Jian et al. \cite{trust} (Cronbach's $\alpha = 0.91$). 

\item \textbf{Competence} was measured with a 5-point scale with 5 items on Perceived intelligence, adapted from the Godspeed questionnaire \cite{anthropomorphism} (Cronbach's $\alpha = 0.77$).

\item \textbf{Likeability} was measured with a 5-point scale with 5 items on Likeability, adapted from the Godspeed questionnaire \cite{anthropomorphism} (Cronbach's $\alpha = 0.88$). 

\item \textbf{Social presence} was measured with a 7-point scale with four items on Engagement, six items on Social Richness and three items on Perceptual Realism adapted from the Temple Presence Inventory (TPI) by Lombard and Ditton \cite{social-presence} (Cronbach's $\alpha = 0.87$). 
\end{itemize}

\subsubsection{\bf Sentiment Analysis of Written Responses}
Sentiment analysis for understanding how the participants felt about the embodiments (negative to positive) was performed on their written responses on the post-survey subjective questions on trust, competence, likeability and socially engagment. We used Microsoft Azure Text Analytics API \cite{microsoft-azure} for the analysis, which provided the score between 0 and 1. The final sentiment score per participant was computed by averaging the scores for all the measures.

\subsection{\bf Behavioral Data and Measures} 

\subsubsection{\bf Economic exchange game for trust} 
The procedure for our economic game was taken from the work by DeSteno et al. \cite{Desteno} which measured trustworthiness of novel partners (human-human and human-robot)  based on the Give-Some Game \cite{Dawes}, a variation of the prisoner's dilemma which allows a wider range of behaviors. In our case, the game is played between the participant and each embodied agent. Each player is given 4 tokens that the player can choose to give to the other player or keep.  Each token worth \$1 if they keep it, but worth \$2 if given to the partner. The maximum social gain occurs when each partner gives all 4 tokens to the other partner (a payoff of \$8 for each partner) resulting in a net social score of 16. The minimum social score occurs when both partners choose to keep all the tokens (a payoff of 4 for each participant) resulting in a net social score of 4. The maximum individual (selfish) payoff occurs when a partner gives no tokens and whose partner gives all four (a payoff of \$12 for the receiver and \$0 for the giver). Thus the game offers a range of social (trustworthy) behavior and also selfish (untrustworthy) behavior.

From the economic exchange game played by the participants, we calculated the \textbf{social payoff}, the net increase in the money as a product of social behaviour, as the sum of tokens given by the participant ($\gamma$) and the tokens the participant predicted to receive ($\rho$). We use the social payoff and normalized by the maximum social payoff to calculate the \textbf{trustworthiness} ($\tau$) score.

\begin{equation}
\tau = \frac{\gamma + \rho}{\gamma_{\text{max}} + \rho_{\text{max}}}
\end{equation}
\newline
The scores were averaged across the games played by the participant with each of the 3 embodied agents to compute the final trustworthiness score of the participant.

%% file: results.tex
\subsection{Effect of Information migration}
An independent t-test was run on the data with a 95\% confidence interval (CI) for the mean difference. We found a significant effect of information migration on Trust (subjective), Trustworthiness (behavioral), Competence, Likeability but not on Social Presence and participants' sentiment. (Figure ~\ref{fig:effect-information}).

\textbf{1. Trust} (subjective) and Trustworthiness (behavioral) scores were found to be significantly higher when the information was migrated, $.668\pm.098$ (Trust) and $.497\pm.209$ (Trustworthiness), than when the information was not migrated, $.611\pm.103$ (Trust) and $.329\pm.168$ (Trustworthiness), with $t$(70)=-2.42, $p$=.018 (Trust) and $t$(70)=-3.73, $p$=.0003 (Trustworthiness).

\textbf{2. Competence} score was significantly higher when the information was migrated ($.721\pm.127$) than when the information was not migrated ($.647\pm.128$) with t(70) = -2.46, p = .016.

\textbf{3. Likeability} score was significantly higher when the information was migrated ($.777\pm.116$) than when the information was not migrated ($.710\pm.135$) with t(70) = -2.26, p = .027. 

\textbf{4. Social Presence} presence score was higher when the information was migrated ($.582\pm.116$) than when the information was not migrated ($.563\pm.124$) with t(70) = -.668, p = .506.

The overall sentiment score of the participants' was positive when the information was migrated ($.603\pm.191$) than when the information was not migrated ($.529\pm.195$) with t(70) = -1.61, p = .112. The effect size (Cohen's d) from the independent t-test on Information migration were Trust (.571), Trustworthiness (.879), Competence (.580), Likeability (.534), Social Presence (.158)

\subsection{Effect of Identity migration}
An independent t-test was run on the data with a 95\% confidence interval (CI) for the mean difference. We found a significant effect of information migration on Trust (subjective), Competence, Social Presence and participants' sentiment, but not on Trustworthiness (behavioral) and Likeability. (Figure ~\ref{fig:effect-identity}).

\textbf{1. Trust} (subjective) score was found significantly higher when the identity was migrated $.666\pm.097$ than when the identity was not migrated $.613\pm.104$ with t(70) = -2.24, p = .028; trustworthiness (behavioral) score was also higher (not statistically significant) when the identity was migrated $.459\pm.227$ than when the identity was not migrated $.369\pm.176$ with t(70) = -1.87, p = .065.

\textbf{2. Competence} score was significantly higher  when the identity was migrated ($.719\pm.109$) than when the identity was not migrated ($.649\pm.146$) with t(70) = -2.33, p = .023.

\textbf{3. Likeability} score was higher (not statistically significant) when the identity was migrated ($.770\pm.128$) than when identity was not migrated ($.717\pm.126$) with t(70) = -1.77, p = .081.

\textbf{4. Social Presence} score was significantly higher when the identity was migrated ($.613\pm.120$) than when the identity was not migrated ($.530\pm.104$) with t(70) = -3.14, p = .002.

The overall sentiment score was significantly positive when the identity was migrated ($.623\pm.183$) than when the identity was not migrated ($.508\pm.193$) with t(70) = -2.61, p = .011. The effect size (Cohen's d) from the independent t-test on Information migration were Trust (.528), Trustworthiness (.441), Competence (.549), Likeability (.418), Social Presence (.741).

\subsection{Effect of Information migration and Identity migration across conditions}
A one-way ANOVA, Fisher's test, across 4 conditions on each measure, was run on the data. We chose to use one-way ANOVA with conditions as independent variable than two-way ANOVA with information migration and identity migration as independent variables because we were interested in comparing the measures across 4 conditions.

\textbf{1. Trust}
Trust (subjective) and Trustworthiness(behavioral) scores had a statistically significant effect across conditions, F(3,68)=5.52, p=0.002 (Trust) and F(3,68)=10.2, p=0.00001 (Trustworthiness). Tukey's HSD test for pair-wise comparison across conditions showed that the trust and trustworthiness in condition (INF+,ID+) was significantly greater than (INF+,ID-) with p=.012 (Trust) and p=.002 (Trustworthiness), (INF-,ID+) with p=.009 (Trust) and p=.00002 (Trustworthiness) and (INF-,ID-) with p=.005 (Trust) and p=.0003 (Trustworthiness). (Figure ~\ref{fig:trust-across-conditions})

\textbf{2. Competence}
We found statistically significant effect across conditions on competence, F(3,68)=4.14, p=0.009. Tukey's HSD test for pair-wise comparison across conditions showed that the competence in condition (INF+,ID+) was significantly greater than (INF-,ID-) with p=.004. (Figure ~\ref{fig:competence-across-conditions}).

\textbf{3. Likeability}
We found statistically significant effect across conditions on likeability, F(3,68)=3.24, p=0.027. Tukey's HSD test for pair-wise comparison across conditions showed that the likeability in condition (INF+,ID+) was significantly greater than (INF-,ID-) with p=.027. (Figure ~\ref{fig:likeability-across-conditions}).

\textbf{4. Social Presence}
We found statistically significant effect across conditions on social presence, F(3,68)=3.39, p=0.023. Tukey's HSD test for pair wise comparison across conditions showed that the social presence in condition (INF+,ID+) was significantly greater than (INF-,ID-) with p=.043. (Figure ~\ref{fig:sp-across-conditions}).

The overall sentiment score in condition (INF+,ID+)=($.687\pm.134$) was significantly positive than (INF-,ID-) = ($.501\pm.184$) with p=.018.

%% file: discussion.tex
\textbf{H1} was partially supported. We found evidence to support that participants reported significantly higher trust, trustworthiness, competence and likeability across the agent embodiments when the information was migrated than not migrated. 

\textbf{H2} was partially supported. We found evidence to support that participants reported significantly higher trust, competence, likeability and social presence across the agent embodiments when the identity of the AI agent was migrated than not migrated. 

When the information was migrated but the identity of the AI agent was not migrated, the participants did not like the fact that the other agents knew about their conversation from previous agent. P51 said \textit{"I did not trust the agents well because they seemed to share all of the information about me, and I did not want to disclose more."}. Also, P39 said \textit{"... especially after the receptionist agent knew what I told Alexa, I no longer trusted Alexa"}. Alternatively, when the identity was migrated but the information of the AI agent was not migrated, the participants did not perceive it to be their own agent. P9 said \textit{"it was not that engaging because it didn't feel that my agent was there"}.

\textbf{H3} Participants perception was reported highest on all the measures in the condition when both information and identity of the AI agent were migrated. This finding is noteworthy because it suggests that both the elements of migration of an AI agent are important to have a significant effect on user's perception. This was corroborated by the comments by the participants in this condition. Participant P21 said  \textit{``I think it remembers that I am anxious about the interview. That means it cares about me and makes it different from, for example, a coffee machine.''} Another participant, P34, said, \textit{``Being familiar with Alexa, allowed me to trust Receptionist and TV agent''}.

%% file: conclusion-and-future-work.tex
We presented a system that enables an AI agent to migrate its persona and information to its next embodiment, thereby providing continuity of task collaboration and context across embodiments. In order to validate the efficacy of the system, we conducted a study to investigate the elements of migration, information and identity migration, on users' perception on trust, competence, likability and social presence across embodiments. We found that the users' perceptions were reported the most positive across all measures when both information and identity of the AI agent was migrated across devices. 

Since the scope of our research question is on the user's perception of information and identity migration across embodiments, we analyzed the experience as a whole using averaging. Analyzing each embodiment effect separately yielded similar trends as the average across embodiments. We also acknowledge that the familiarity with the brand of agents (Amazon/Google) could have had an effect on users' perceptions. However, we observe that even in cases of strong brand presence (Google logo on Smart TV), the general trend of user perception remained the same as in agents with no strong brand presence.

